\documentclass[twocolumn,twoside,slac_two]{revtex4}
\usepackage{graphicx}
\usepackage{fancyhdr}
\usepackage{graphics}
\usepackage{epstopdf}
\usepackage{textpos}
\begin{document}

\title{\centering Fully hadronic ttbar cross section measurement with ATLAS detector}
\author{
\centering
\begin{center}
 C.Bertella, on behalf of the ATLAS Collaboration
\end{center}}
\affiliation{\centering CPPM, Marseille 13288 France}
\begin{abstract}

The top quark pair production cross section in the fully hadronic
   final state is characterized by a six jet topology, two of which
   could be identified as originating from a $b$-quark using ATLAS
   $b$-tagging algorithms. Compared to other decay channels, this final
   state presents an advantageous larger branching ratio; on the other
   hand it suffers from a very large QCD multi-jet background,
   generally difficult to estimate from Monte Carlo simulation and
   therefore evaluated using data-driven techniques.
 
  The analysis  is performed using $36 \mathrm {pb^{-1}}$ of pp collisions produced at the LHC with a center-of-mass energy
of  7 TeV. The observed upper limit is set at 261 pb at 95\% confidence level, where the expected Standard Model cross-section for the $t\bar{t}$ process is $165^{+11}_{-16}$ pb.
   In the future, when the LHC luminosity increases, it is essential, to efficiently trigger on these fully hadronic $t\bar{t}$ events, to use dedicated triggers.
    An overview of the analysis for $t\bar{t}$ production cross section
   measurement in the fully hadronic final state and the
   state-of-the-art of the $b$-jet trigger performance estimation are
   presented in this contribution.
     \end{abstract}

\maketitle
\thispagestyle{fancy}
\section{Introduction}

 The top quark, discovered at Fermilab in 1995, completed the three generation structure of the Standard Model, SM. The top quark is
distinguished by its large mass, ($m \sim 172~$GeV). 
Precision measurements in the top quark sector will shed light on the electroweak symmetry breaking mechanism and indirectly on the Higgs mechanism of elementary particle mass generation.
In particular, the measurement of the top quark pairs ($t\bar{t}$) production cross section is an important test of QCD perturbative calculations as well as an estimation of one of the major background sources for several new physics signatures. 
The full hadronic $t\bar{t}$ is important background for several analyses beyond the Standard Model: producton of new particle(s) decaying to many hadronic jets in association with missing trasverse momentum, predicted by SUSY; search for the Higgs boson in the all hadronic ($bb+$jets) final state, such as associated production with vector boson or top pair.

In proton-proton collisions, 
top-antitop pairs are created when a parton from each colliding proton interact through the strong force.
The production mechanisms at the LHC are the gluon-gluon fusion (85\%) and $q\bar{q}$ annihilation (15\%)

Within the SM, the top quark decays into a $W$ boson and a $b$-quark almost ${100\%}$ of the time. The W boson subsequently decays into either a pair of quarks or a lepton-neutrino pair. In the $t\bar{t}$ production in the fully hadronic final state both $W$s decay hadronically.

The experimental signature of fully hadronic $t\bar{t}$ is characterized by a nominal  six-jet topology with $b$-jets.
This channel has the advantage of a large BR,  top pair decay in full hadronic signature in (${44\%}$) of the case, although it suffers from a large QCD multijet background.\\

The ATLAS \cite{ATLAS} detector at the LHC covers nearly the entire solid angle around the collision point. It consists of an inner tracking detector divided in three independent subsystems immersed in a 2T magnetic field generated by the central
solenoid that reconstructs charged particle trajectories and measures their momentum,  an electromagnetic calorimeter that identifies and measures the electrons and photons and a hadronic calorimeter that identifies jets formed by the hadronization of quarks and muon spectrometer that  identifies muons and measures their deflections in the magnetic field produced by a toroid magnet system (4T).

\vspace{-0.37cm}

\section{Analysis method}

\vspace{-0.01cm}

During the 2010, ATLAS recorded ${\sim 36\ \mathrm{pb^{-1}}}$ of p-p collisions at center of mass energy of 7 TeV.\\
The data sample has been collected with unprescaled multijet triggers, requesting at least four jets with pseudorapidity\footnote[1]{In the right-handed ATLAS coordinate systems, the pseudorapidity $\eta$ is defined as $\eta = - \ln[\tan(\theta/2)]$, where the polar angle $\theta$ is measured with respect to the LHC beamline.
Transverse energy is defined $E_T = E \sin\theta$.}, $|\eta|$, less than 3.2 and transverse energy, $E_T$, great than 30 GeV; the event selection requirement is of at least four jets with $E_T>60~$~ GeV \cite{koei}. The trigger efficiency for a jet with its $E_T = 60$ GeV is 90\%. 
The identification of jets originating from $b$-quarks ($b$-tagging) is performed using a secondary vertex-based tagger algorithm which reconstructs the inclusive vertex formed by the decay products of bottom hadron.\\

\section{Background modeling}

\begin{figure}
\includegraphics[scale=0.23]{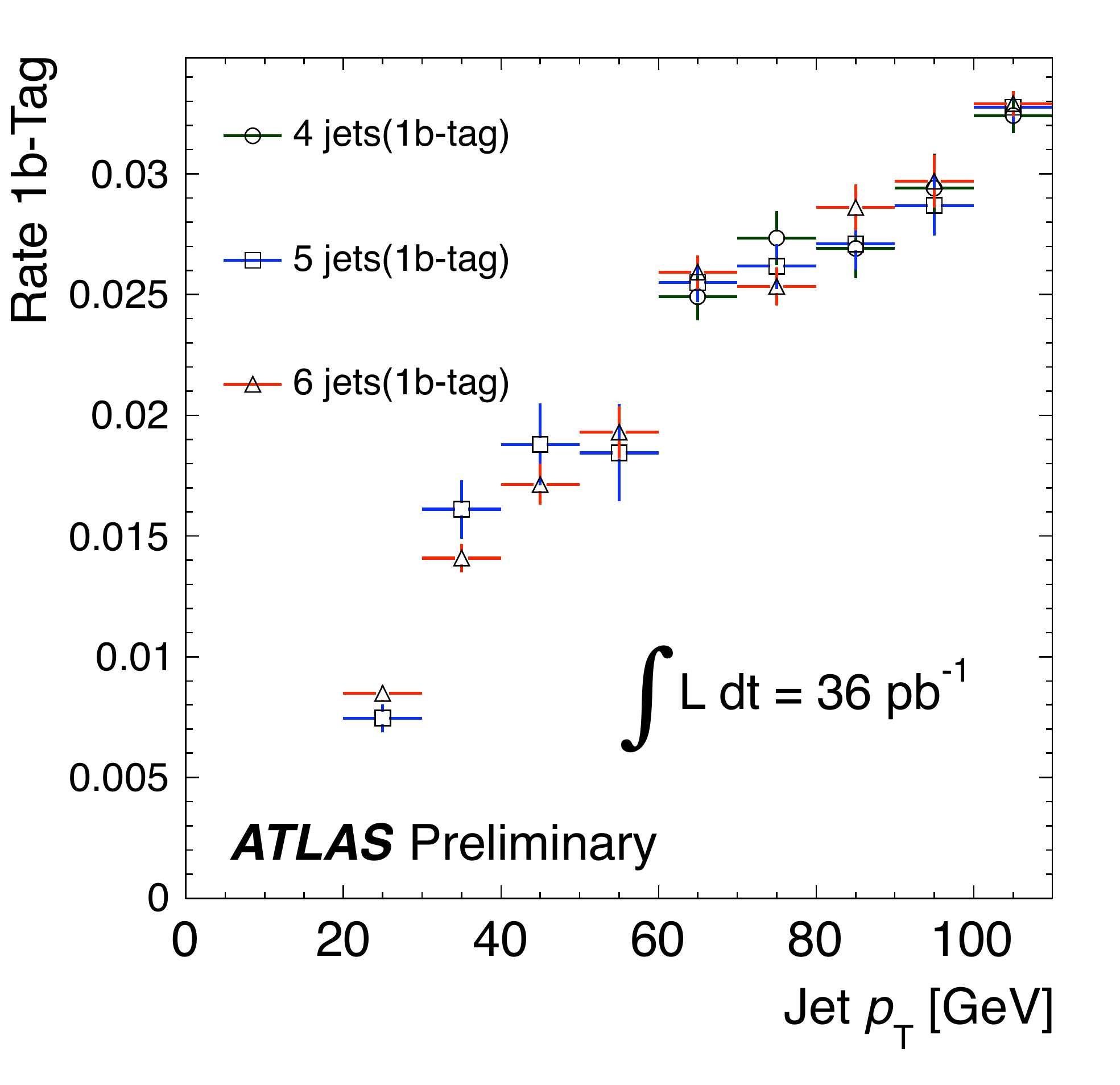}
\caption{Background tag rate function as a function of jet $p_T$ from data \cite{koei}.}\label{TRF}
\end{figure}

The QCD multijet background shape is modelled by a data-driven method applying the tag-rate function. 
The signal is defined as events with 6 or more jets with 2 $b$-tagged jets.
The baseline tag rate function (TRF) to be used for the backgroud modelling in the signal region are derived from the 5 jets control bin with 2 $b$-tags, as seen in Fig. \ref{TRF}.
The tag rates are evaluted separately for 1 $b$- and 2 $b$-tag events, because the event topology in the 2 $b$-tag case is different due to gluon splitting process ($g\rightarrow b\bar{b}$). 
The background tag rate for a jet (TR) is defined as:

\begin{equation}
TR_{nb_{bin},nj_{bin}} = \frac{njet_{tag}}{njet_{all}} \times \frac{nj_{bin}}{nb_{bin} \times {_nj_{bin}}C_{nb_{bin}}}
\end{equation}
The $TR_{nb_{bin},nj_{bin}}$ function is calculated in each jet and b-tag multiplicity bin separately. The variables $njet_{all}$ and $njet_{tag}$ are respectively the total number of jets and the number of $b$-tagged jets, the symbol C stands for the binomial coefficient $(_{n}C_{r} = \frac{n!}{(n-r)!r!})$ for jets to be selected as $b$-tagged jets in the event.
 
\begin{figure}
\includegraphics[scale=0.23]{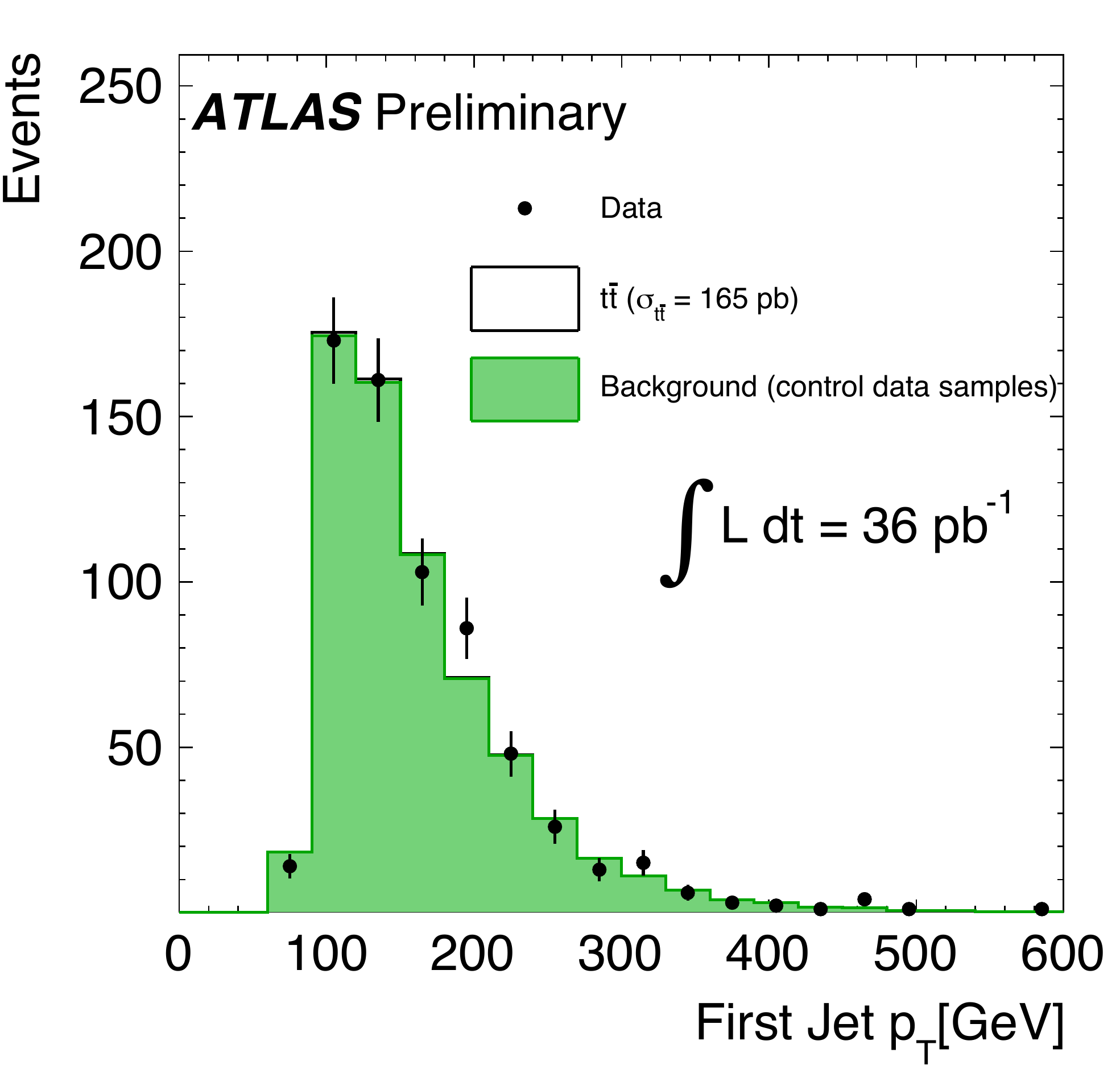}
\caption{Kinematic distributions for 4-jet 2 $b$-tag events (Control Region) \cite{koei}. }\label{kin}
\end{figure}

The variables ${njet_{tag}}/{nb_{bin}}$ and ${njet_{all}}/{nj_{bin}}$ correspond to the number of tagged events in $nb_{bin}$ and  $nj_{bin}$ and the number of pretag events in ${nj_{bin}}$, respectively. The coefficient $_{n}C_{r}$ is a factor to make this probability on a per-jet basis. $TR_{nb_{bin},nj_{bin}}$ is to be applied to
a jet to obtain an event weight. The TRF is parametrized as a function of the jet $p_T$ and $\eta$ to take into account a possible dependence on those quantities.  
The weight for an event with 6 jet (2 $b$-tag) is obtained applying as follows the TRF:

\begin{equation}
w(6j,2b)= \sum_{k=1}^{nj_{bin}} TR^k_{njb_{bin},5j} \times \frac{{_{nj_{bin}}}C_{nb_{bin}}}{nj_{bin}}
\end{equation}
After applying TRF kinematic distribution are compared in control region, 4 jet 2 $b$-tag events, see fig. \ref{kin}.\\
\section{Fitting procedure}

The background normalization, for the cross section measurement, is extracted from a fit to the mass $\chi^2$ variable
\begin{equation}
\chi^2 = \sum_{i=1}^2 (\frac{m_{jjb}^{(i)}-m_{top}}{\sigma_{top}})^2 +(\frac{m_{jj}^{(i)}-m_{W}}{\sigma_{W}})^2
\end{equation}
Only the leading 6 jets are considered in $\chi^2$ computation for $N_{jet}\ge~6$ events. 
In the case of 2 $b$-tag events (signal) there are 6 combination, and we select the combination with the minimum $\chi^2$ to build the final $\chi^2$ distribution.

 \subsection{$t\bar{t}$ cross section measurement analysis with 2010 data}
\begin{figure}[h!]
\includegraphics[scale=0.24]{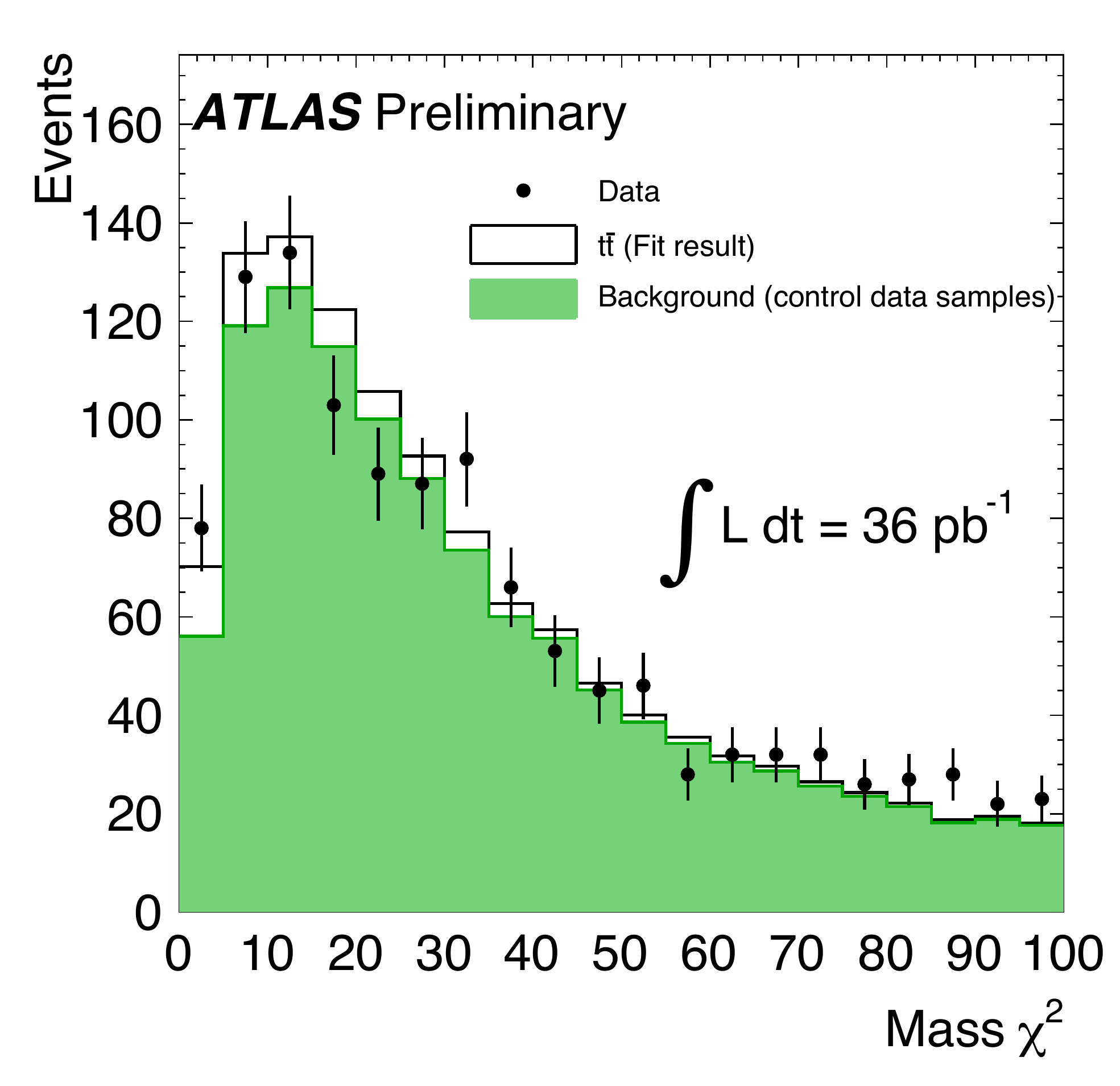}\caption{Fit result of minimum mass $\chi^2$ distribution in 6-jet with 2 $b$-tag events (Signal Region) \cite{koei}.}\label{fig:fit1}
\end{figure}

The final mass $\chi^2$ distribution is fitted with signal and background template. Figure \ref{fig:fit1} shows the fit result of the minimum mass $\chi^2$ distribution.
The cross section $\sigma_{t\bar{t}}$ is obtained by
\begin{equation}
\sigma_{t\bar{t}} = \frac{N_{obs} \times f_S}{\epsilon \times \int L dt}
\end{equation}
where $f_s$ is the signal fraction of 6.4\% from the fit; the factor $\epsilon$ of 1.8\% includes signal acceptance and branching franction.

\begin{table}[]
\begin{center}
\caption{Fit results for events with at least 6-jets and 2 $b$-tags}
\begin{tabular}{|c|c|}
\hline \textbf{Source} & \textbf{Number of events} \\
\hline
 Background & 1097.0 \\
$t\bar{t}$ & 75.0$\pm $46.5 (stat.)\\
\hline
Data ($N_{obs}$) & 1172\\
\hline
\end{tabular}
\end{center}
\end{table}

The fitted cross section with 36 $\mathrm{pb^{-1}}$ is

 $\sigma_{pp \rightarrow t\bar{t}} = 118 \ \pm \ 73  (\mathrm{ stat.} ) \ \pm\  48  (\mathrm{ syst.} ) \ \pm \  4  (\mathrm{ lumi.} ) \mathrm{pb}$.
 The significance of the fitted value is $1.6 \ \sigma$ whereas the  expected sensitivity was $2.2 \ \sigma$.
The observed one-sided upper limit at 95\% confidence level is $\sigma_{t\bar{t}}<~261\mathrm{ pb}$. 
The total systematic uncertainty on the $t\bar{t}$ cross-section is 41\%. The most important  systematic sources considered are Jet energy scale uncertainty(JES), Trigger efficiency, b-tagging, Background modeling, Initial \& Final State Radiation (ISR \& FSR).
Table \ref{syst} shows a summary of the most important individual contributions.
\begin{table}[h!]
\begin{center}
\caption{Summary of  the most important individual contributions on the systematic uncertainties on the fitted cross-section value}
\begin{tabular}{|c|c|}
\hline 
\textbf{Source} & $\Delta \sigma/ \sigma$ \\
\hline
JES & 17\%\\
Trigger & 10\%\\
b-tagging & 29\%\\
Background modeling & 7\%\\
ISR/FSR & 16 \%\\
\hline
\end{tabular}\label{syst}
\end{center}
\end{table}

\vspace{-0.8cm}

\section{ Outlook: $b$-jet trigger}
The ATLAS trigger and data acquisition system is based on three levels.
Trigger levels must provide a rejection to reduce the 40 MHz bunch-crossing rate to an output of about few hundred Hz.
The level 1 is hardware based, uses the calorimeter and muon spectrometer with coarse granularity;  the level 2 is software-based, exploits regions of interest identified by the level 1 and accesses data from all sub detectors with full granularity; the Event Filter (EF) runs offline-quality software-based algorithms.

\subsection{Online $b$-tagging}
The identification of jets stemming from the hadronization of $b$-quarks is made possible by the\\relatively long lifetime of hadrons containing $b$-quarks (lifetime of the order of $1.5\ ps$ corresponding to $c\tau\approx  450\mu m$). This allows the identification of $b$-jets from the one containing only lighter quarks. 
Given the high instantaneous luminosity the LHC will deliver in the next year, $b$-tagging at HLT is a possibility for collecting 
$t\bar{t}$ in the full hadronic final state with an acceptable data taking rate.
\vspace{+0.2cm}

\subsection{Future analysis}
\begin{figure}[h!]
\includegraphics[scale=0.27]{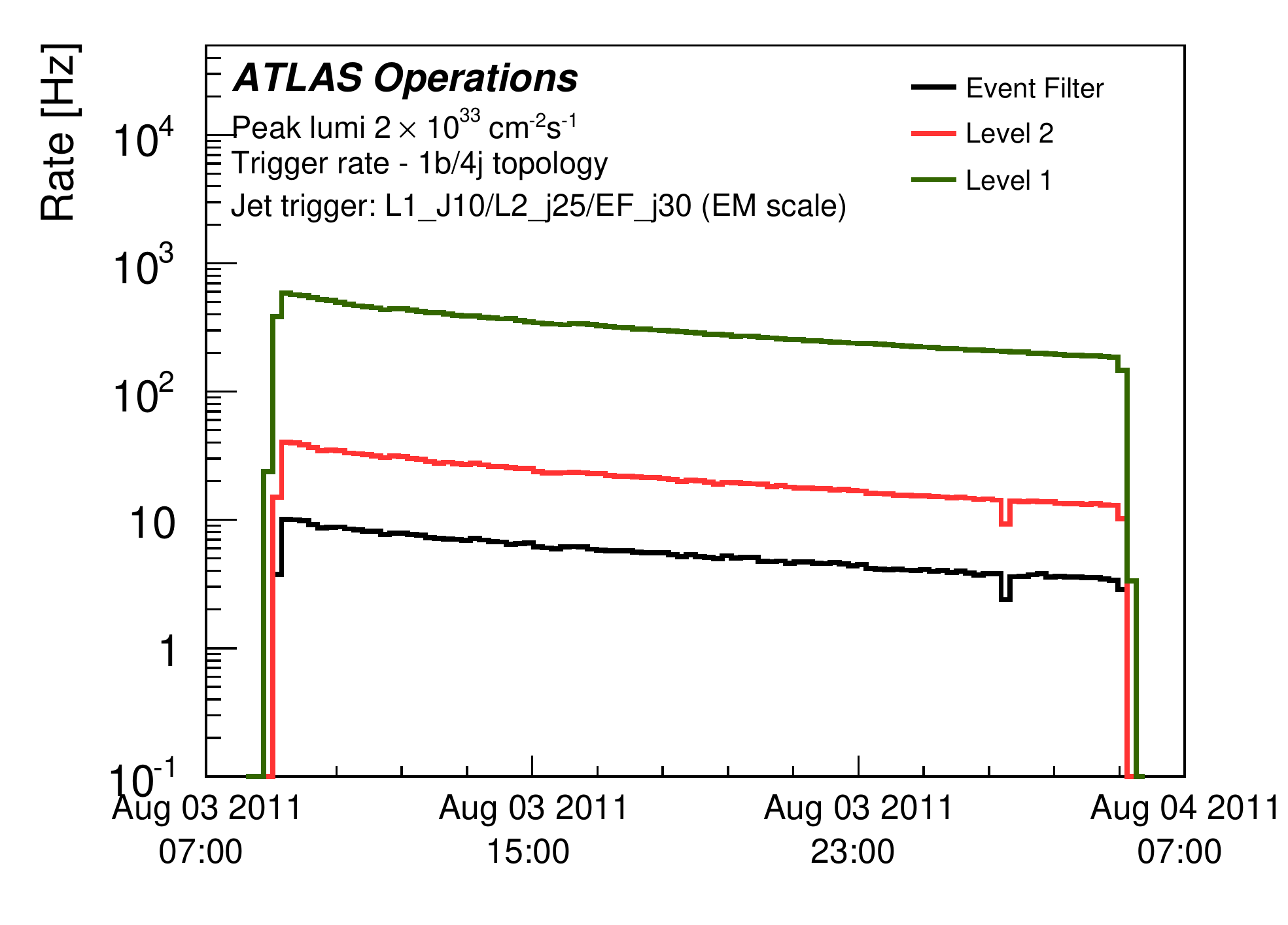}\caption{Trigger rate for 1b/4j topology. LVL1, LVL2 and EF rate of a b-jet trigger requiring at least four jets in the event and at least one $b$-tagged jet \cite{btrig}.} \label{b1}
\end{figure}

At the high instantaneous luminosity foreseen for LHC data taking, most multi-jet trigger will be prescaled unless their threshold and jet multiplicity are constantly increased to keep the trigger rate under control. For this reason ATLAS has put in place a combination of multijet and b-jet trigger to efficiency select events with final states containing several $b$-jets.\\
The $b$-jet trigger for hadronic top requires four EF-jets with $E_T>~30~$GeV at electromagnetic scale and 1 $b$-jet with $E_T>~10~$GeV at EM scale and tight instance for the $b$-tagging criteria, as seen in Fig. \ref{b1}.
The signal efficiency is $\sim40\%$.

\vspace{-0.8cm}

\section{Conclusions}
The analysis is performed using 36 $\mathrm {pb^{-1}}$ of pp collisions produced at the LHC with a center-of-mass energy of $\sqrt{s} = 7$ TeV and recorded with the ATLAS detector. A 95\% confidence level limit is set at 261 pb, compatible with the expected Standard Model cross-section of $165^{+11}_{-16}$ pb.

\vspace{-0.5cm}

\bigskip
\bibliography{basename of .bib file}

\end{document}